\documentclass{svjour3}  
\usepackage[table,xcdraw]{xcolor}
\usepackage{float}
\usepackage{amssymb}
\usepackage{amsmath}
\usepackage{cite}
\usepackage{supertabular,booktabs}
\usepackage{supertabular}
\usepackage{lipsum}                     
\usepackage{xargs}                      % Use more than one optional parameter in a new commands
\usepackage{listings}
\usepackage{color}
\usepackage{todonotes}
\definecolor{dkgreen}{rgb}{0,0.6,0}
\definecolor{gray}{rgb}{0.5,0.5,0.5}
\definecolor{mauve}{rgb}{0.58,0,0.82}
\usepackage[utf8]{inputenc}
\usepackage[colorlinks]{hyperref}
\usepackage{amsxtra}  
\usepackage{amstext}
\usepackage{amssymb}
\usepackage{graphicx}
\usepackage{amsfonts}
\usepackage{multicol}
\usepackage{setspace}

\usepackage{tikz}
\usetikzlibrary{external}
\usetikzlibrary{shapes,arrows,chains}
\usetikzlibrary[calc]
\tikzset{external/force remake}
\tikzstyle{line} = [draw, -latex']

\definecolor{proliferating_green}{HTML}{4af04a}
\definecolor{complex_blue}{HTML}{4a98ff}
\definecolor{red_dead}{HTML}{ff3d81}
\definecolor{necrotic_gray}{HTML}{808080}
\definecolor{alien_blue}{HTML}{000000}
\definecolor{ecm_blue}{HTML}{6cb8eb}
\definecolor{nutrient_orange}{HTML}{f5b44c}

\begin{document}

%\begin{frontmatter}
\title{Evaluation of Entropy and Fractal Dimension as Biomarkers for Tumor Growth and Treatment Response using Cellular Automata}

\author{Juan Uriel Legaria-Peña \and
        Félix Sánchez-Morales \and
        Yuriria Cortés-Poza\textsuperscript{(1)}}

\institute{(1) Y. Cortes-Poza \at
              IIMAS, Unidad Académica de Yucatán, Universidad Nacional Autónoma de México (UNAM), Yuc., México.
             \email{yuriria.cortes@iimas.unam.mx}           
           \and
           J.U. Legaria-Peña  \email{walup@ciencias.unam.mx}
            \and
           \ \\
           F. Sánchez-Morales 
                \email{felixsm@ciencias.unam.mx}
}

\

\date{Received: date / Accepted: date}

\maketitle

\begin{abstract}
     Cell-based models provide a helpful approach for simulating complex systems that exhibit adaptive, resilient qualities, such as cancer. Their focus on individual cell interactions makes them a particularly appropriate strategy to study the effects of cancer therapies, which often are designed to disrupt single-cell dynamics. In this work, we also propose them as viable methods for studying the time evolution of cancer imaging biomarkers (IBM). We propose a cellular automata model for tumor growth and three different therapies: chemotherapy, radiotherapy, and immunotherapy, following well-established modeling procedures documented in the literature. The model generates a sequence of tumor images, from which time series of two biomarkers: entropy and fractal dimension, is obtained. Our model shows that the fractal dimension increased faster at the onset of cancer cell dissemination, while entropy was more responsive to changes induced in the tumor by the different therapy modalities. These observations suggest that the prognostic value of the proposed biomarkers could vary considerably with time. Thus, it is important to assess their use at different stages of cancer and for different imaging modalities. Another observation derived from the results was that both biomarkers varied slowly when the applied therapy attacked cancer cells in a scattered fashion along the automatons' area, leaving multiple independent clusters of cells at the end of the treatment. Thus, patterns of change of simulated biomarkers time series could reflect on essential qualities of the spatial action of a given cancer intervention.   
\end{abstract}


%\begin{keyword}
%Avascular tumor modeling \sep Complex Systems \sep Cellular Automata \sep Imaging Biomarkers \sep Shannon Entropy \sep Fractal Dimension
%\end{keyword}

%\end{frontmatter}

\newpage

\newpage
%Biomarker Research Springer
%Introduction
\section{Introduction}

\textit{Cancer} is a disease that can develop in multi-cellular organisms, where uncontrolled cell proliferation occurs due to genetic mutations. It currently stands as one of the main threats to human health, accounting for approximately 10 million yearly deaths according to data provided by the world health organization \cite{WHO}. As a result, substantial efforts have been made to develop treatments against the disease, mainly focusing on disrupting the genetic and functional processes that molecular biology has identified to be related to its development. However, these treatment efforts have only partially succeeded since cancer tissue's heterogenous and adaptive qualities make it resilient and capable of producing various survival strategies. This difficulty has produced a paradigm shift in recent years. Cancer is now viewed as a complex adaptive system, and research efforts have been increasingly focused on understanding how underlying interactions between cancer cells produce the macroscopic structures and versatile dynamics that characterize the pathology \cite{cancer_complex_system}.
Multiple mathematical and computational techniques have been proposed to study how the dynamics of cancer emerge from local microscopic conditions and interactions. One particularly convenient approach is cell-based computational modeling, where the properties of cells are stated as individual discrete entities that can interact with one another \cite{cell_based_modelling}. Models of this kind have been applied to research multiple aspects of cancer, such as the emergence of calcifications in breast cancer, vessel morphology in angiogenesis, and adaptive phenomena such as the go or grow effect observed in Glioblastoma Multiforme \cite{ductal_carcinoma,angiogenesis_tip_overtaking, go_or_grow}. 

A widely applied subclass of cell-based models is cellular automata (CA). CA methods involve lattices, where each site can be occupied by a single cell. At each simulation step, rules for how a cell will respond to its environment are applied, and local actions such as cell division, motility, or death are carried through. These models have found critical applications in studying cancer therapies' effect on tissue since most therapeutical strategies focus on disrupting individual cell properties or capabilities \cite{cas_cancer_therapy}. 

Assessment of therapy and disease progression often requires the use of biomarkers. A \textit{biomarker} can be defined as an indicator with a utility for characterizing normal physiological processes, diseases, or responses to treatments \cite{biomarker_defs}. In the management of cancer, the introduction of novel imaging techniques such as positron emission tomography (PET), single‐photon emission computed tomography (SPECT), and magnetic resonance imaging (MRI) has allowed the proposal of novel imaging biomarkers (IBM), most of them focused on tumor size characteristics. The use of tumor size as a cancer indicator is mostly prioritized because the gold standard for tumor response evaluation (RECIST) uses mass longest diameters as a reference to classify therapy effects \cite{recist_guidelines}. Some studies, however, have highlighted the need to complement such criteria with other biomarkers, particularly for types of tumors such as hepatocellular carcinoma, lung cancer, prostate cancer, brain glioma, and lymphoma, where other considerations such as necrosis or cell clustering might become relevant \cite{beyond_recist}.

In this work, we propose a Cellular Automata model for tumor growth and three cancer interventions: radiotherapy, chemotherapy, and immunotherapy. Synthetic images generated with the model are used to evaluate the time series of two prospective biomarkers: Shannon's entropy and fractal dimension, as the tumor develops and when each of the modeled treatments is administered. These state measures have been shown to change as the complexity and configuration of systems are altered. 

Since exposing a patient to multiple imaging procedures is not clinically viable, using computational models to study biomarkers and their evolution over time becomes a handy tool. Our model produces high-resolution images of the evolution of the tumor during the growing stage and treatment. The proposed biomarkers of these images are analyzed. 

% Estructura del articulo

The work presented here has the following structure: Section 2 provides a brief review; Section 3 details the different models implemented; Section 4 presents our results; Section 5 discusses the ideas obtained from these results; we conclude our work in Section 6.

\section{Background}
The methods used in this study derive from several disciplines, such as mathematics, biology, medicine, physics, and computer science. This section provides a comprehensive theoretical base that could be useful to researchers from any of the abovementioned fields.

\subsection{Cancer as a pathology}
Cancer is a disease where an abnormally high proliferation of cells takes place \cite{cancer_definition}. This high division rate is often accompanied by other adaptive, resilient traits and hallmarks, such as immune evasion and a marked glycolytic metabolism \cite{cancer_hallmarks}. 

The disease tends to progress in two identifiable stages. In the first one, an avascular tumor mass grows until reaching a final stationary size. At this point, the tumor cells struggle to gather nutrients by passive diffusion, and a process of vessel recruitment known as angiogenesis starts to develop. This phenomenon constitutes the start of the second cancer stage and has been linked to crucial signaling chemicals such as vascular endothelial growth factors (VEGF) \cite{angiogenesis}. In addition, cancer cells can travel through the bloodstream in their latest stages, invading distant body parts. This extensive dissemination of the disease receives the name of metastasis \cite{cancer_stages}. 

\begin{figure}[!htbp]
    \centering
    \includegraphics[width = 0.65\linewidth]{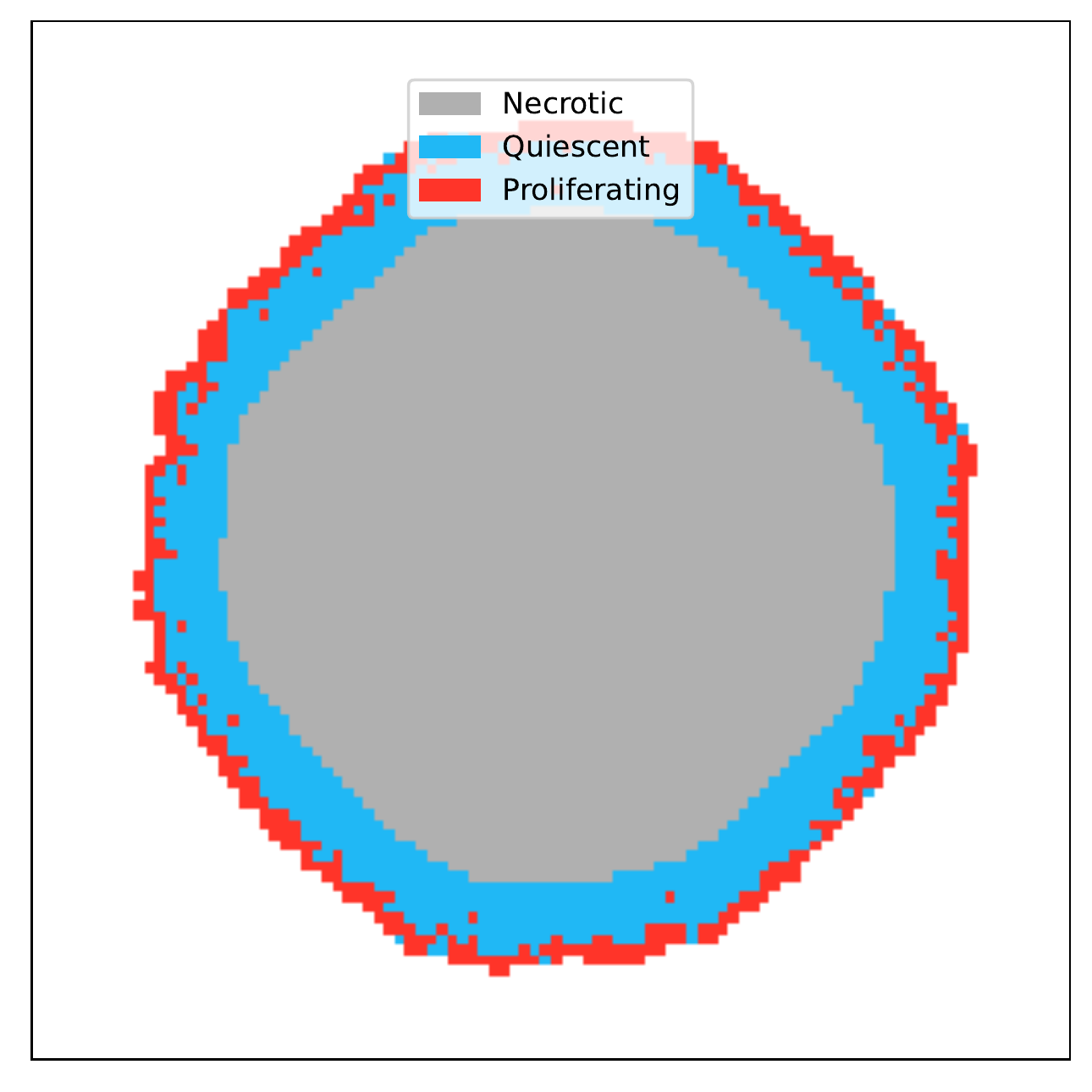}
    \caption{Layer organization of cells in a tumor. The image was generated with the tumor model developed in this work.}
    \label{fig:tumor_layers}
\end{figure}

The model presented in this work will focus solely on avascular tumor growth. As already mentioned, in this phase, the tumor develops into a stable mass, where nutrients will be most readily available for those cells lying on the outer surface of the conglomerate. Differences in nutrient absorption results in a particular arrangement of cells: those at the core of the tumor die, leaving necrotic scar tissue; cells at intermediate portions of the tumor stay in a quiescent dormant state that is less metabolically demanding, and actively proliferating cells reside on the outer layers of the tumor. This morphological organization of tumor cells has been schematized in Figure \ref{fig:tumor_layers}.

\subsection{Cancer therapies}\label{sec:cancer_therapies}

Several known therapies are used to slow the progression of the disease and, in some cases, to revert it.
The ones modeled in this work are radiotherapy, chemotherapy, and immunotherapy. From a systems perspective, these interventions can be thought of as control methods devised to alter the dynamics of cancer cells and drive the system out of its diseased state.

Some cancer therapies are designed to act more efficiently at certain stages of the cell cycle. The cell cycle is a periodic progression of events that a cell goes through to achieve division. It comprehends four stages named G1, S, G2, and M. G1 is a phase where the cell grows and increases its number of organelles, the S phase is where the replication of DNA takes place, G2 constitutes a stage where cells produce all the material needed for division, and finally, M phase or mitosis is the process of nuclear division. 

It is worth mentioning that selecting the treatment to apply follows established medical guidelines that consider factors such as cancer stage, location, and other relevant indicators. These protocols have the purpose of optimizing cancer outcomes for most patients, and they are constantly revised and updated based on new findings.

The mechanisms of radiotherapy, chemotherapy, and immunotherapy and the biological effects they exert on cancer cells are the following:

\begin{itemize}
    \item \textit{Radiotherapy}: In radiotherapy, high doses of gamma radiation are applied to the affected tissue, resulting in a cascade of effects damaging proliferating cells. Its action proceeds in two stages. In the first one, which takes a short time, ionizing biochemical material crucial for survival might kill some cells. Furthermore, in the long term, a phenomenon known as radiolysis takes hold, where radicals of water molecules generated by irradiation might produce compounds such as oxygen peroxide, which are toxic to the cell, and may induce its death. Radiation tends to act more effectively on cells in advanced stages of their cell cycle \cite{radiotherapy_cell_cycle}.
    
    \item \textit{Chemotherapy}: In Chemotherapy, drugs that damage the cancer cell's genetic material are administered to the patient. The effects are dependent on local drug concentration and the cycle stage. Cells in the S-phase are the ones most likely to die as a result of the treatment \cite{chemotherapy_cell_cycle}. 
    
    \item \textit{Immunotherapy}: Immunotherapy comprises interventions destined to increase immune system detection and targeting of cancer cells. The one that we use in our model consists of training T-cells ex-vivo for cancer antigen recognition. This therapy is administered gradually to the patient. 
\end{itemize}

\subsection{Cellular Automata fundamentals}

This work will simulate avascular tumor growth and treatment using Cellular Automata. This model was first introduced by John von Neumann in his studies of self-replicating machines and has been widely used for studying how collectively organized structures can emerge in lattices of individual cells \cite{neuman_automata, automata_complex_systems}. In CA, each grid cell changes its state at every simulation step based on the state of its close neighbors and other local conditions.

\begin{figure}[!htbp]
    \centering
    \includegraphics[width = 0.75\linewidth]{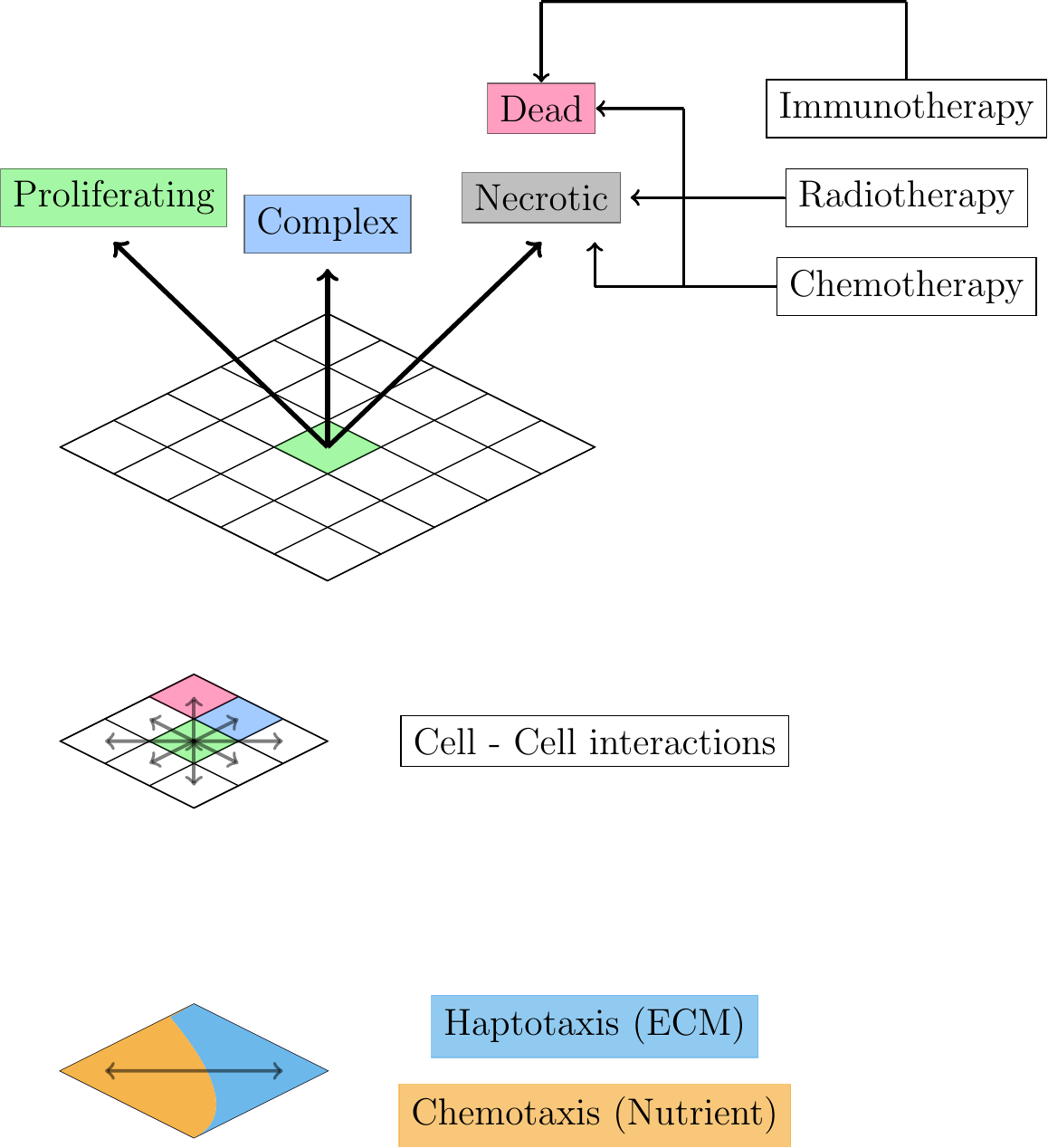}
    \caption{Elements considered in the proposed model for cancer growth and therapy.}
    \label{fig:automata_summary}
\end{figure}

The Cellular automaton implemented in this work to simulate tumor growth and cancer therapies is presented in Figure \ref{fig:automata_summary}. For tumor growth, a rectangular grid will be used, and the state of each space will denote the type of biological cell currently occupying that location. The state values considered are proliferating, where cancer cells can divide depending on environmental conditions, complexes formed by entrapment of cancer cells by the immune system; and two types of death: necrotic, where scar tissue is left, and dead state, where the cell is cleanly disposed, and can later be replaced by healthy tissue (vacant spaces in the automaton).

Changes in a cell's state will depend mainly on cell-cell interactions with its neighbors. Rules for the possible state transitions will be specified in section \ref{sec:cell_cell_interactions}. 

Aside from communication with adjacent cells, the model considers two critical factors that regulate the directional proliferation of cancer cells: haptotaxis, mainly induced by inward constraining contact forces exerted by extracellular matrix (ECM), and chemotaxis, an effect where nutrient concentration gradients drive tumor cell propagation outwards. These two effects can oppose each other, and local differences can determine the tumor's final morphology and extension of the tumor \cite{chemotaxis_and_haptotaxis}. In the automata model, the proliferation and necrosis of cells at a given grid location will be conditioned on ECM and Nutrient concentrations. Also, the dynamical degradation of ECM by cancer cells and spatial diffusion of nutrients will be simulated. 

Therapies in the model act mainly by stochastic induction of cancer cell death: concretely, radiotherapy and chemotherapy will produce either clean death or necrosis, while immunotherapy will always dispose of cells leaving no scar tissue. 

\subsection{Biomarkers for complexity: Shannon's entropy and fractal dimension}

Recent advances in medical imaging techniques have opened new possibilities for the types of analysis that can be conducted to characterize cancer processes accurately. In particular, Shannon's entropy has been used successfully to segment and detect tumors in MRI images \cite{entropy_ml_1, entropy_ml_2}. Some studies have also pointed to entropy as an appropriate indicator to characterize cancer in CT images, given that a correct calibration for confounding factors is established \cite{entropy_ct}. 

Shannon's entropy in an image can be computed using its normalized grey-scale values histogram $P_{i}$, where $i$ indexes the bins used to group intensity values. The equation that defines it is the following:

\begin{equation}
    S = \sum_{i = 0}^{N} P_{i}\log{\left(\frac{1}{P_{i}}\right)}
\end{equation}

In addition to entropy, another complex-systems-related quantity associated with cancer processes is the fractal dimension. Some studies have found that fractal dimension in images obtained from histological studies can be a helpful indicator in detecting cancer cell proliferation \cite{frac_dim_tumor_1, frac_dim_tumor_2}. 

The fractal dimension is closely related to the notion of self-similarity, that is, the presence of structures that repeat themselves at multiple scales in geometrical patterns or representations of data. Self-similarity is a recurrent property exhibited by biological systems, being present in neural networks, the immune system and most relevant to this work in tumor cell arrangements \cite{fractal_bio_1, fractal_bio_2, fractal_bio_3}.

A commonly applied method to compute fractal dimension is box-counting, where the number $M$ of boxes of size $\epsilon$ required to fill the analyzed geometrical structure is obtained. An estimation for fractal dimension can be derived by applying the following equation:

\begin{equation}
    D = \frac{M}{\left(\frac{1}{\epsilon}\right)}
\end{equation}

In practice, $D$ is rarely computed for just one value of $\epsilon$. Instead, a standard statistically robust method consists of fitting a linear regression model to the pairs $(\epsilon, M)$, and then the fractal dimension can be found as the slope of the adjusted line.

\section{Proposed model}
Details of the proposed cellular automata for tumor growth and treatment will be given in this section. In addition, our computational model implementation can be consulted at \cite{github_link}.

\subsection{Tumor Growth}\label{sec:tumor_growth_model}

As already schematized in Figure \ref{fig:automata_summary}, the base tumor growth model comprehends three essential elements to establish: rules for how the state of individual cells will change (for instance, from healthy vacant spaces to proliferating cancer cells), haptotaxis qualities of the modeled tissue, namely ECM degradation by the expanding tumor, and chemotactic and necrotic effects due to changes in nutrient concentration. 

\subsubsection{Cell - cell interactions and state transitions}\label{sec:cell_cell_interactions}

The proposed cell-cell automaton interactions are based on the work of Shahmoradi et al. in \cite{shahmoradi}. The proposed derivation, however, includes chemotactic-driven growth and necrosis of cells due to nutrient insufficiency. 

\begin{figure}[h]
    \centering
    \includegraphics[width = \linewidth]{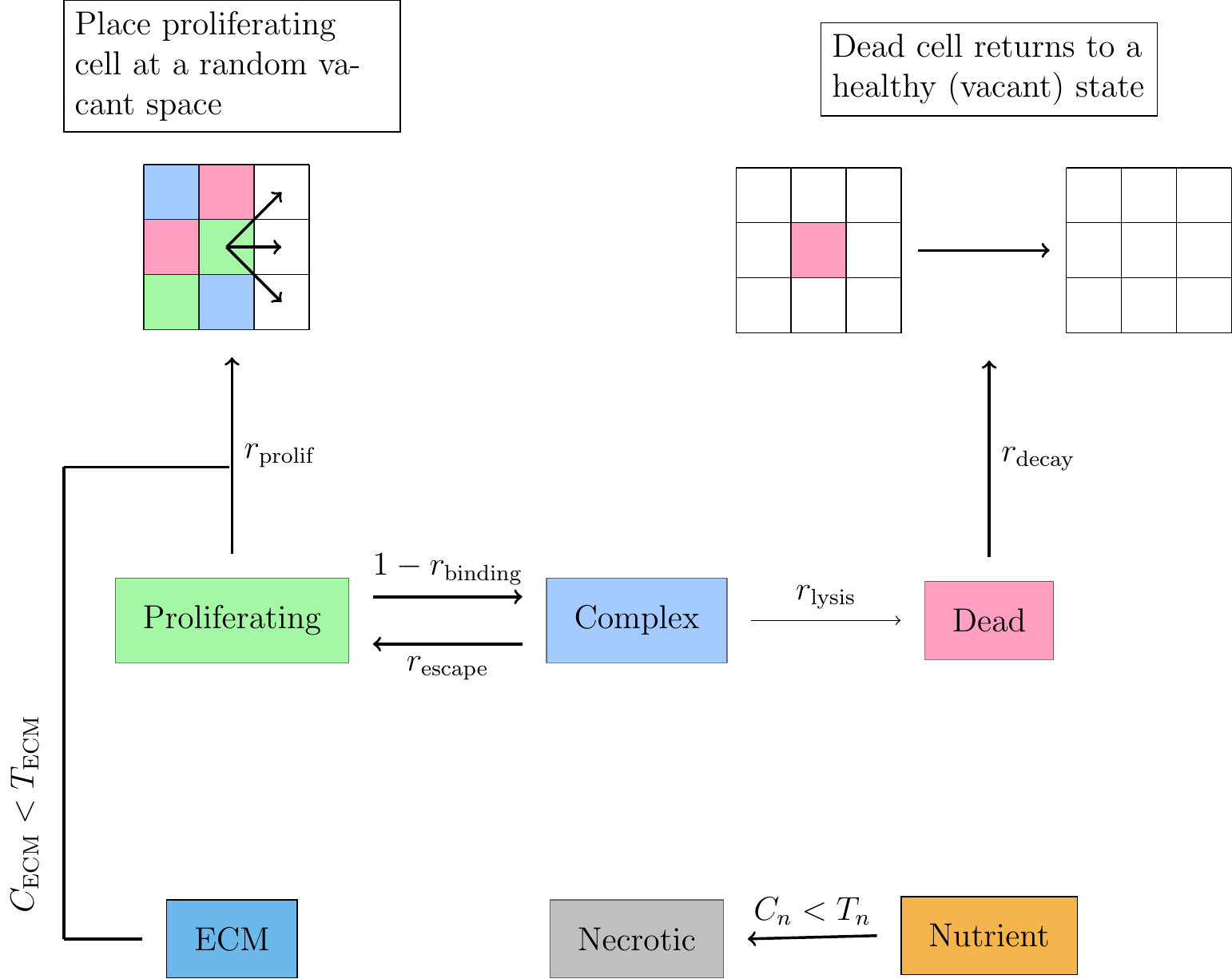}
    \caption{State transition diagram of the automaton's cells}
    \label{fig:cell_state_transitions}
\end{figure}

A diagram of the possible transitions between cell states is summarized in Figure \ref{fig:cell_state_transitions}.

Proliferating cells divide at every step of the automaton's evolution with a probability $r_{\text{prolif}}$. Cell division involves randomly placing a new proliferating cell at any available (healthy) surrounding spaces, and it is also conditioned by the concentration of extracellular matrix $C_{\text{ECM}}$ at the location where the system attempts to place a new cell. Namely, if this concentration is more significant than a threshold $T_{ECM}$, the placement of the sprung daughter cell will be negated. 

The division probability $r_{\text{prolif}}$ is limited by the number of proliferating cells in the tumor. If $N_{\text{prolif}}$ is the current number of proliferating cells and $N_{\text{prolif}, \text{max}}$ is a carrying limit for such cells, then the following equation is applied:

\begin{equation}
    r_{\text{prolif}} = r_{\text{prolif}, 0}\left(1 - \frac{N_{\text{prolif}}}{N_{\text{prolif}, \text{max}}}\right)
\end{equation}

where $r_{\text{prolif}, 0}$ is the initial maximum division probability.

When a proliferating cell does not achieve division, it can form a complex with immune cells with a probability $1 - r_\text{binding}$. Antibody immune complexes can undergo one of two faiths, they could either end up killing the attacked cancer cell with probability $r_{\text{lysis}}$ at every step, or the entrapped cell could escape and return to its previous proliferating state with probability $r_{escape}$. 

Finally, dead cells will be able to return to healthy vacant spaces with probability $r_{\text{decay}}$ at every step since they were killed by clean immune catalysis. The secondary type of death considered, necrosis, can be reached whenever the concentration of nutrient $C_{n}$ at a cell's position drops below a threshold $T_{n}$. Cells in this state can not be occupied and remain effectively vanished for the rest of the simulation.

\subsubsection{Haptotatic extracellular matrix (ECM)}
%ECM diffusion 
Haptotaxis involves a simulation of the process where cancer cells progressively degrade ECM as they come in contact with it. The evolution rule used to update the concentration of ECM is given by the following equation:

\begin{equation}
    C_{\text{ECM}}^{\text{new}}(i,j) = C_{\text{ECM}}(i,j) - e_{c}I(i,j)C_{\text{ECM}}(i,j)
    \label{eq:ecm_concentration_update}
\end{equation}

where $e_{c}$ is a constant and $I(i,j)$ is the number of proliferating cells adjacent to the location $i,j$ of the automaton. 

\subsubsection{Chemotactic nutrient gradient}

Nutrient diffusion, which determines chemotaxis and necrosis, was modeled using the second Fick's law. Thus, the update rule for the nutrient's concentration $C_{n}$ is given by the following equation:

\begin{equation}
    C_{n}^{\text{new}}(i,j) = C_{n}(i,j) + D_{n}\nabla^{2}C_{n}(i,j) - C_{\text{abs},n}
    \label{eq:nutrient_concentration_update}
\end{equation}

where $D_{n}$ is a diffusion constant, $C_{\text{abs},n}$ is an absorption of the nutrient by the cells, and $\nabla^{2}C$ requires computing the discrete laplacian with concentration values at neighboring cells, as stated in Equation \ref{eq:discrete_laplacian}.

\begin{align}
     &\nabla^{2}C_{n}(i,j) = C_{n}(i+1,j) + C_{n}(i-1,j) \notag\\
    &+ C_{n}(i,j+1) + C_{n}(i,j-1) -4C_{n}(i,j)
    \label{eq:discrete_laplacian}
\end{align}

For the cell's nutrient absorption, two values were considered: one $C_{\text{abs, prolif}}$ which is applied when cells have surrounding vacant spaces to proliferate, and a lower value  $C_{\text{abs, qui}}$, used for quiescent cells that have no available neighbor positions where to put new daughter cells. 

\begin{figure}[h]
    \centering
    \includegraphics[width = \linewidth]{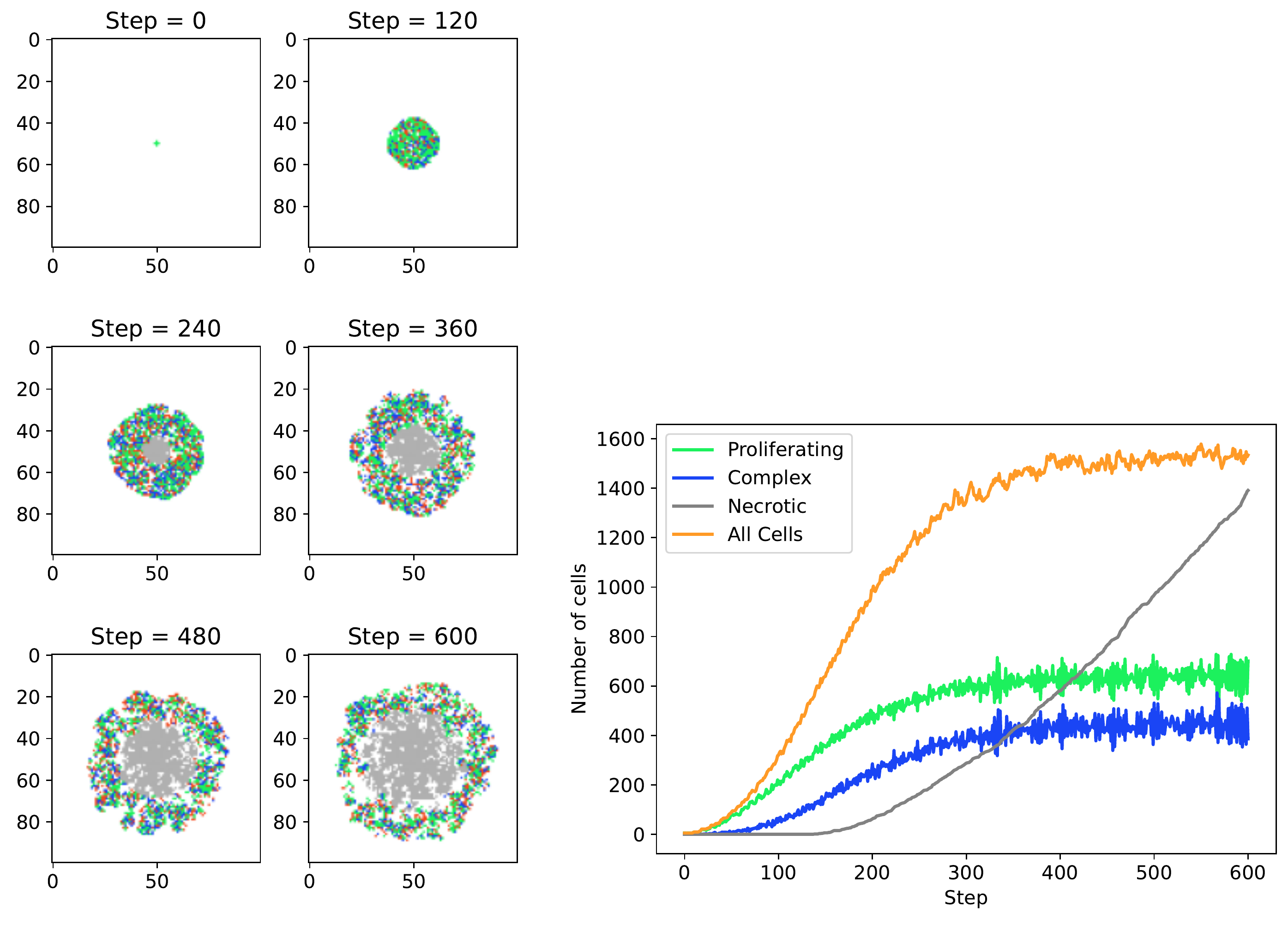}
    \caption{Automaton images and cell counts obtained in a case where no therapy was administered. Evolution took 600 steps and parameters values used were $r_{\text{prolif},0} = 0.85$, $r_{\text{binding}} = 0.1$, $r_{\text{escape}} = 0.5$, $r_{\text{lysis}} = 0.35$, $r_{\text{decay}} = 0.35$, $N_{\text{prolif}, \text{max}} = 1000$, $e_{c} = 0.1$, $D_{n} = 0.05$, $C_{\text{abs}, \text{prolif}} = 0.01$, and $C_{\text{abs}, \text{qui}} = 0.005$}
    \label{fig:automaton_state_evol}
\end{figure}

Figure \ref{fig:automaton_state_evol} shows images of the automaton's state taken at six different time steps and also time series for the counts of the different types of biological cell states considered in the model. 
At every grid location, initial concentrations of ECM are assigned a random value in the interval $(0.8, 1.2)$. 
Cells are initialized with a nutrient concentration of 1; the ones at the boundary are assigned a fixed value of 2.
Finally, we start with four proliferating cells in a cross shape at the center of the grid.

\begin{figure}[h]
    \centering
    \includegraphics[width = \linewidth]{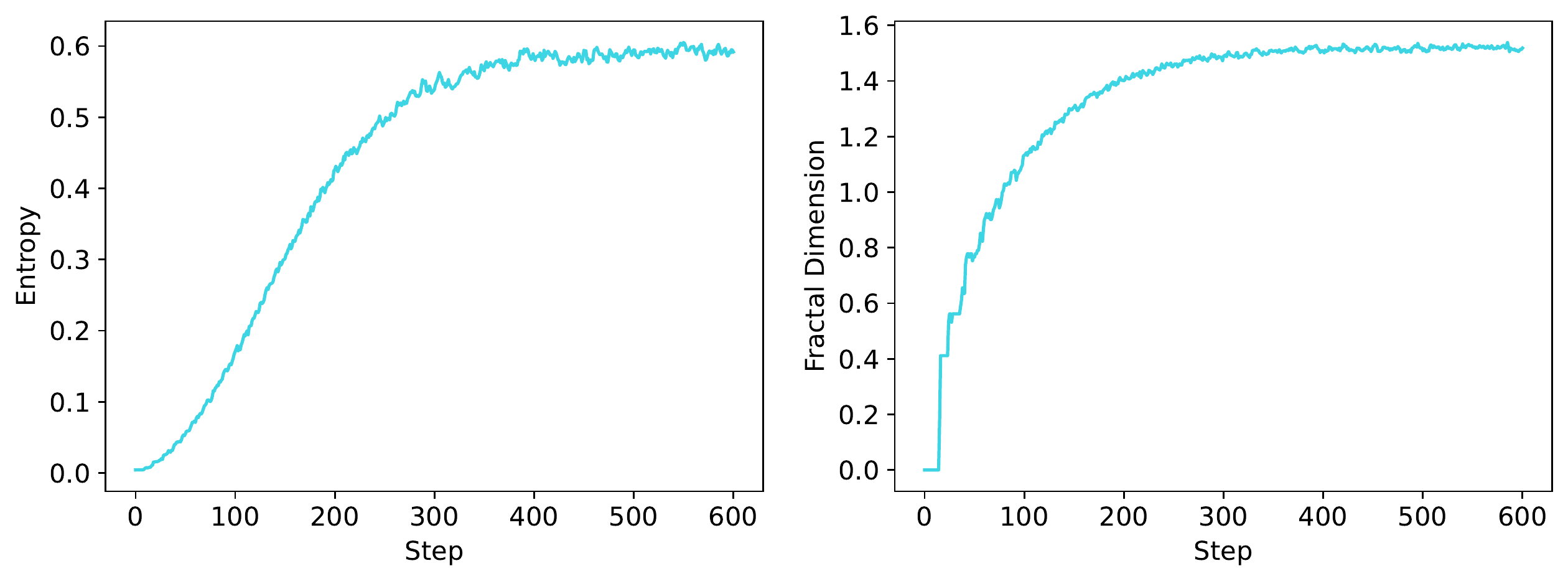}
    \caption{Entropy and fractal dimension time series obtained with the automaton images in a case where no treatment was applied.}
    \label{fig:no_treatment_entropy_frac}
\end{figure}

Time series for entropy and fractal dimension obtained with the simulated images as the tumor grew are shown in Figure \ref{fig:no_treatment_entropy_frac}.

\subsection{Radiotherapy model}

The model for radiotherapy was based on the linear-quadratic cell radiation response model used by Sarah C. Brüningk et al. in \cite{bruning}. 

Radiotherapy delivers a simulated radiation beam at step $n_{0, \text{rad}}$ to the automaton's tissue. The probability that a cell at location $i,j$ is affected by the treatment is then computed using the following response function:

\begin{equation}
    P_{\text{rad}}(i,j) = 1 - \exp{\left(-\gamma\left(\alpha d_{\text{OER}}(i,j) + \beta d_{\text{OER}}^{2}(i,j)\right)\right)}
    \label{eq:probability_target_radiotherapy}
\end{equation}

In the previous expression, $\alpha$ and $\beta$ are constant parameters, weighing the linear and quadratic responses of the tissue. $\gamma$ is related to the cell cycle stage, explained in section \ref{sec:cancer_therapies}, and can be obtained using Equation \ref{eq:gamma_factor_calc}.

\begin{equation}
    \gamma = \gamma_{0}\times 1.5^{(n - s_{0})\%4}
    \label{eq:gamma_factor_calc}
\end{equation}

where $\gamma_{0}$ is a constant and $s_{0}$ is the step at which the cell in question was first placed in a vacant space. 

The term $d_{\text{OER}}$ appearing in Equation \ref{eq:probability_target_radiotherapy} is called the oxygen status of the cell, and it is a factor dependent on the nutrient concentration $C_{n}$ (in this case interpreted as oxygen saturation) and on the applied gamma beam dose $d$. Its calculation can be carried out using Equations \ref{eq:oxygen_status_1} and \ref{eq:oxygen_status_2}.

\begin{align}
    &d_{OER} = \frac{d}{\text{OER}}\label{eq:oxygen_status_1}\\
    &\text{OER} = \begin{cases}
    1 & C_{n}(i,j) \geq T_{n, \text{rad}}\\
    1 -\frac{C_{n}(i,j)}{T_{n, \text{rad}}} & C_{n}(i,j) < T_{n, \text{rad}} \label{eq:oxygen_status_2}
    \end{cases}
\end{align}

In Equation \ref{eq:oxygen_status_2}, $T_{n, rad}$ is a low boundary threshold for oxygen concentration, and those cells where $C_{n}$ drops below such value will have the greatest probability of being targeted by radiotherapy.

Cells selected to be targeted by the treatment by the previous response function will undergo either death or necrosis with probability $P_{\text{rad}, \text{0}}$ for the first $\tau_{\text{rad, delay}}$ steps after applying the radiation beam, and with a probability $P_{\text{rad},f}$ for the rest of the simulation. Here a ratio $f_{\text{nec}}$ of deaths that will be of the necrotic type is fixed at the start of the automaton's evolution.

\begin{figure}[h]
    \centering
    \includegraphics[width = \linewidth]{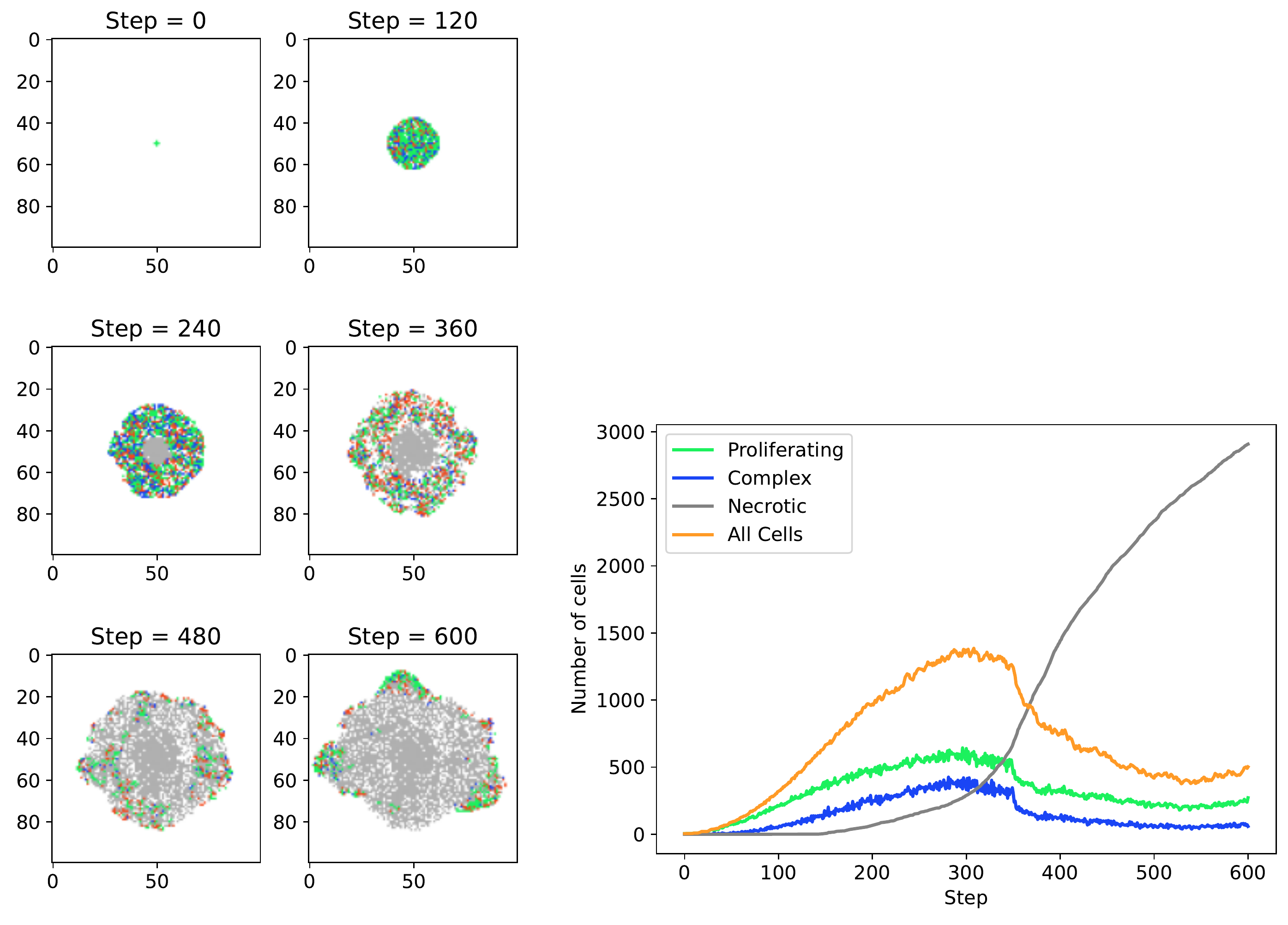}
    \caption{Images and cell count time series for the evolved tumor when radiotherapy was administered halfway through its growth. Parameters used for the radiotherapy treatment were: $\gamma_{0} = 0.05$, $\alpha = 0.1$, $\beta = 0.05$, $d = 1$, $T_{n,\text{rad}} = 0.35$, $\tau_{\text{rad,delay}} = 50$, $P_{\text{rad},0} = 0.02$ and $P_{\text{rad},f} = 0.5$}
    \label{fig:radio_automaton_state_evol}
\end{figure}

Figure \ref{fig:radio_automaton_state_evol} shows images recorded at six different instants of the automaton's evolution and treatment. In this test, the tumor was allowed to grow for the first 300 steps, and then one single application of radiation was delivered to the tissue. The right side of the figure also shows the number of cells counted at every simulation instant. 

\begin{figure}[h]
    \centering
    \includegraphics[width = \linewidth]{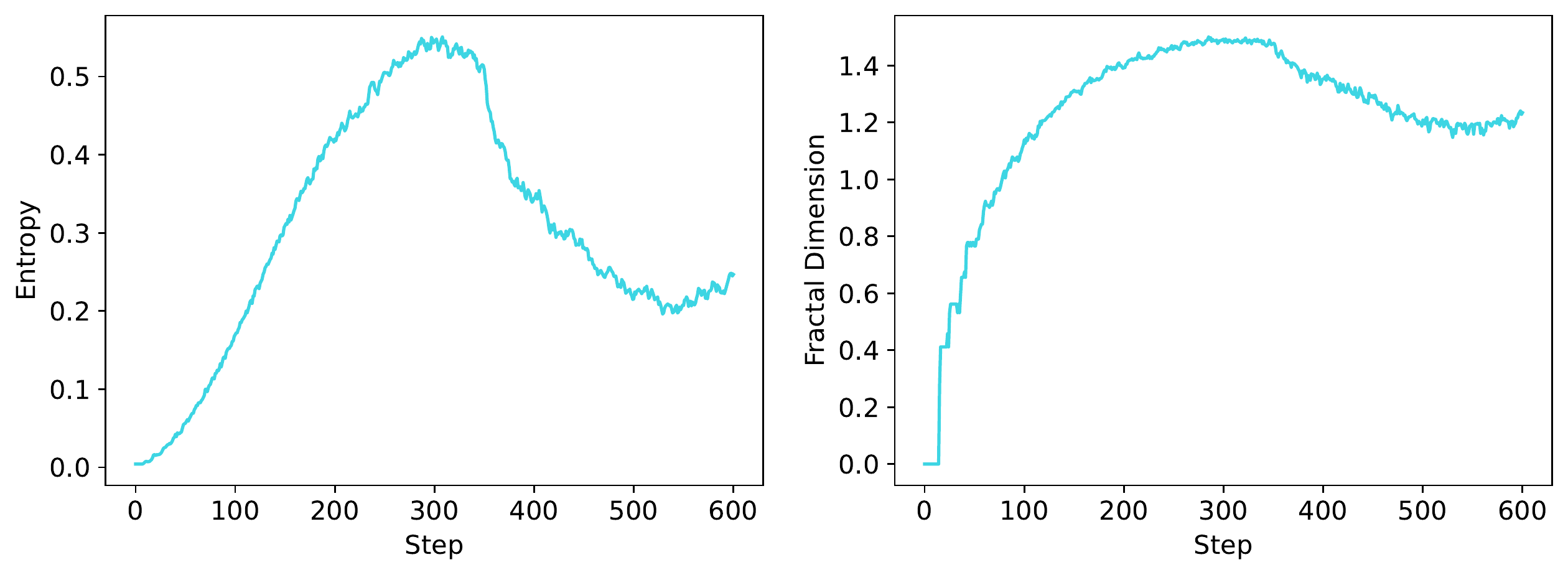}
    \caption{Entropy and fractal dimension time series obtained in a case where a beam of radiation was administered midway through the tumor development (step 300).}
    \label{fig:rad_treatment_entropy_frac}
\end{figure}

Corresponding time series for entropy and fractal dimension are shown in Figure \ref{fig:rad_treatment_entropy_frac}.

\subsection{Chemotherapy model}

The chemotherapy model applies a cell response function similar to the one provided for the radiotherapy scenario. In this case, the response model presented by Fateme Pourhansanzade and S.H. Sabzpoushan in \cite{fateme} was used. 

When chemotherapy starts at step $n_{0, \text{chem}}$, a constant drug concentration $C_{0,d}$ is set at the borders of the automaton for a period of $\tau_{\text{chem}}$ steps, simulating the injection of pharmaceuticals. The administered drug will then perfuse the automaton's area for the rest of the simulation following the second Fick's law for diffusion. An update of the medication concentration at position $i,j$ of the grid will thus follow the next rule:

\begin{equation}
    C_{d}^{\text{new}}(i,j) = C_{d}(i,j) + D_{d}\nabla^{2}C_{d}(i,j)
\end{equation}

where $D_{d}$ is the drug's diffusion constant and $\nabla^{2}C_{d}$ is obtained as the discrete laplacian (See Equation \ref{eq:discrete_laplacian}).

Drug concentration $C_{d}$ determines a given cell's response to chemotherapy. The probability of the cell at location $i,j$ dying at step $n$ of the simulation is given by Equation \ref{eq:chemotherapy_probability_death}. Here two types of death are considered: normal death and necrotic, where the necrotic ratio of dead cells by chemotherapy is established at the beginning of the simulation.

\begin{equation}
    P_{\text{chem}} = l_{m}(i,j)\times PK \times \exp{\left(-c_{m}\left(n - \tau_{sim}\right)\right)}
    \label{eq:chemotherapy_probability_death}
\end{equation}

In equation \ref{eq:chemotherapy_probability_death}, $\tau$ is the number of steps left in the simulation after starting chemotherapy, and $PK$ is a pharmacokinetics factor that modulates how quickly the injected drug is metabolized. Furthermore, $c_{m}$ is an attenuation factor associated with live cells in state $m$ (either proliferating or complex in our case), and $l_{m}(i,j)$ is a linear factor related to the medicine concentration $C_{d}(i,j)$. The latter can be computed using the following equation:

\begin{equation}
    l_{m}(i,j) = \frac{k_{m}\times C_{d}(i,j)}{CR_{m}'\times \tau_{sim} + 1}
    \label{eq:chemo_linear_factor}
\end{equation}

In the previous expression, $k_{m}$ is the killing rate of cells in state $m$, and $CR_{m}'$ is a chemical resistance parameter assigned stochastically with relation \ref{eq:stochastic_chem_resistance}.

\begin{equation}
    CR_{m}' = CR_{m}\times \text{random(0,1)}
    \label{eq:stochastic_chem_resistance}
\end{equation}

where $CR_{m}$ is the maximum possible value for chemical resistances and $\text{random(0,1)}$ denotes a floating point value taken in the interval between 0 and 1.

An essential aspect of the chemotherapy model we considered is that the medication can target only cells at the S-phase of their cell cycle. The cell cycle stage is computed within the simulation with Equation \ref{eq:cell_cycle_stage}.

\begin{equation}
    \text{cell stage} = (n - s_{0})\%4    
    \label{eq:cell_cycle_stage}
\end{equation}

Thus, only those cells with cell stage value one will be damaged by radiation. 

Also, we consider that a certain fraction $R_{\text{chemo}}$ of cells will be treatment-resistant and natively unaffected by the administered drugs. These cells have the quality of asymmetric division; that is, they can either produce a new resistant cell with probability $P_{\text{res}, \text{chemo}}$, or their offspring can be non-resistant. In contrast, non-resistant cells will always produce new daughter cells which are non-resistant, i.e., their division will be symmetric.

\begin{figure}[h]
    \centering
    \includegraphics[width = \linewidth]{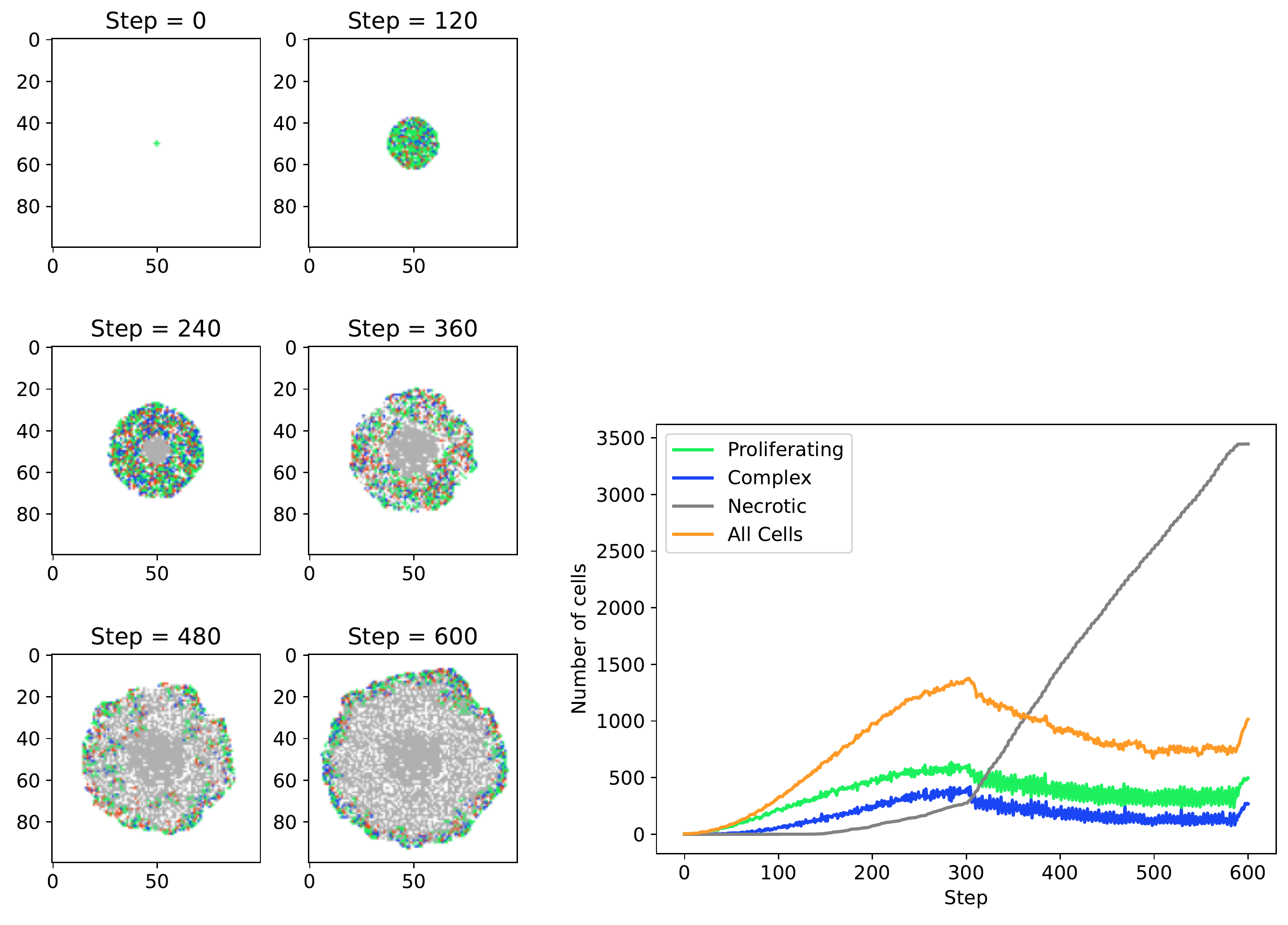}
    \caption{Cell counts and images of an evolved tumor automaton, where chemotherapy treatment is applied at step 300. Value parameters used for this reported test were: $D_{d} = 2$, $PK = 1$, $R_{\text{chemo}} = 0.1$, $\tau_{\text{sym}} = 300$, $\tau_{\text{chem}} = 2$, $C_{d,0} = 1$, $c_{\text{Proliferating}} = 0.5$, $c_{\text{Complex}} = 0.5$, $CR_{\text{Proliferating}} = 0.1$, $CR_{\text{Complex}} = 0.05$, $k_{\text{Proliferating}} = 0.8$, $k_{\text{Complex}} = 0.01$}
    \label{fig:chemo_automaton_state_evol}
\end{figure}

Figure \ref{fig:chemo_automaton_state_evol} shows the results of tumor treatment with chemotherapy in a scenario where drugs are administered in step 300 of the automaton's evolution. The right side of this figure shows cell counts obtained for the different cell states considered. 

\begin{figure}[h]
    \centering
    \includegraphics[width = \linewidth]{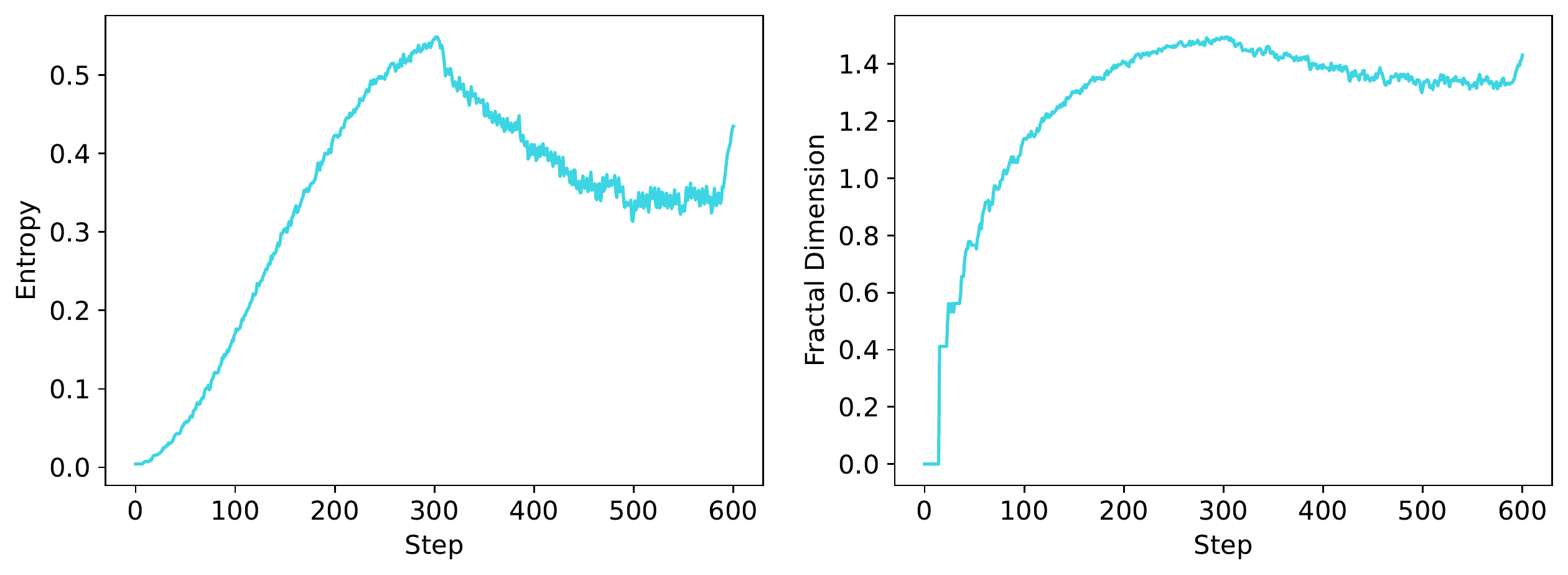}
    \caption{Entropy and fractal dimension time series obtained in a case where chemotherapy was delivered starting at step 300.}
    \label{fig:chemo_entropy_frac}
\end{figure}

Entropy and fractal dimension time series obtained in the chemotherapy case are shown in Figure \ref{fig:chemo_entropy_frac}. 

\subsection{Immunotherapy model}

Immunotherapy was based on the automaton model devised by Shamoradi et al. in \cite{shahmoradi}. It is based on a therapy where T-cells are trained to better detect and dispose of cancer cells. The therapy is gradually delivered to a patient. 

In the automaton, this effect can be imitated by slowly changing the parameters so that the formation of immune complexes (binding of proliferating cells and the immune system) increases. These complexes metabolize the attacked cells more efficiently. 

If $r_{\text{final}}$ and $r_{\text{initial}}$ are the target and initial values for a given tumor growth parameter, and the effects of chemotherapy are distributed over a number $\tau_{\text{immuno}}$ of steps, then the following increment would be added to the parameter value at every simulation step to modify it:

\begin{equation}
    \Delta r = \frac{r_{\text{final}} - r_{\text{initial}}}{\tau_{\text{immuno}}}
\end{equation}

until the number of steps $\tau_{\text{immuno}}$ that the therapy takes is reached. 

\begin{figure}[h]
    \centering
    \includegraphics[width = \linewidth]{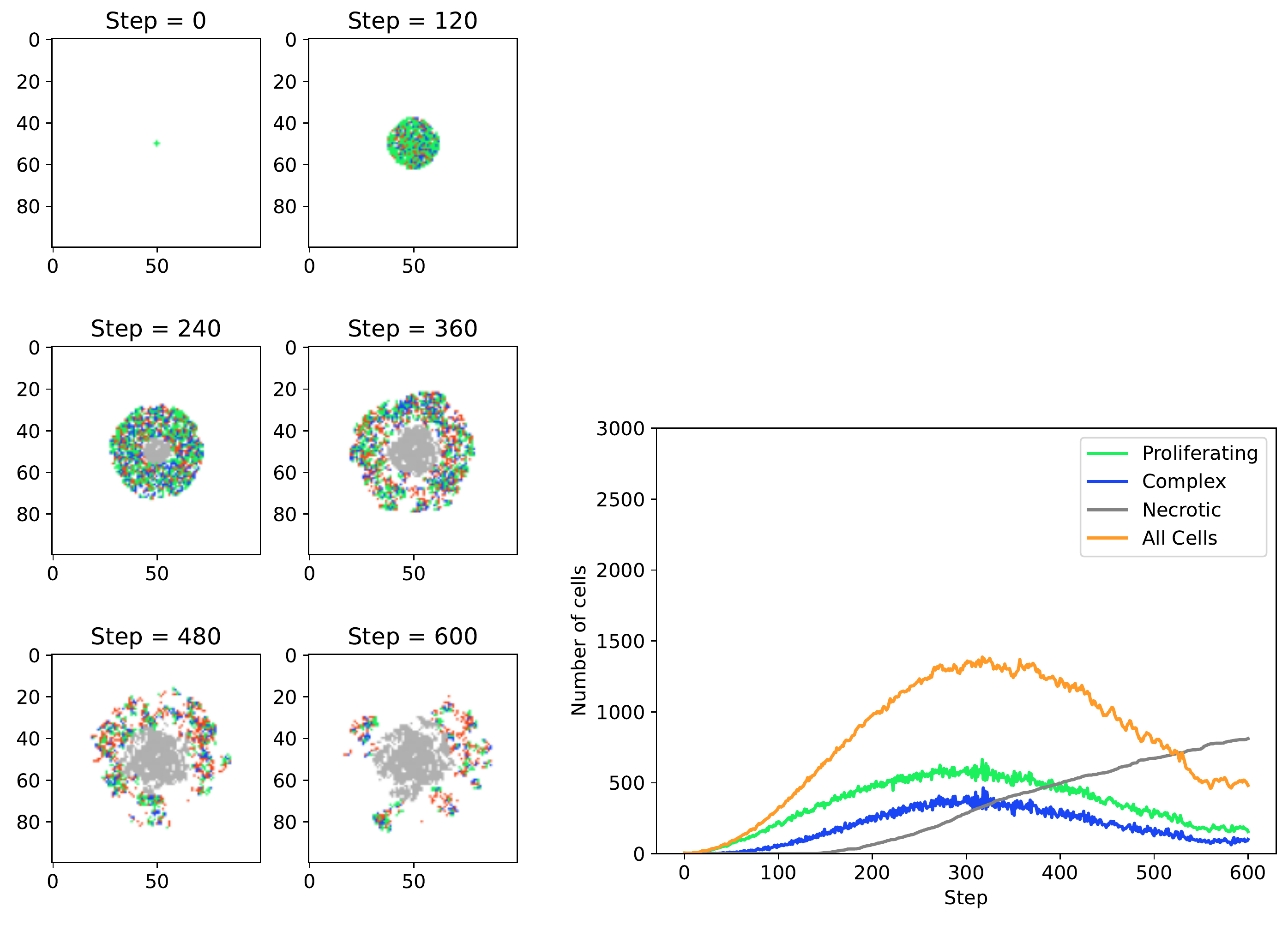}
    \caption{Images of the tumor growth automaton's state and cell counts obtained for a case where immunotherapy was delivered at step 300. Parameters used for the treatment were: $r_{\text{final, prolif}} = 0.65$, $r_{\text{final, binding}} = 0.001$, $r_{\text{final, escape}} = 0.001$, $r_{\text{final,lysis}} = 0.9$, $r_{\text{final,decay}} = 0.35$ and $\tau_{\text{immuno}} = 250$.}
    \label{fig:immuno_automaton_state_evol}
\end{figure}

Figure \ref{fig:immuno_automaton_state_evol} shows the evolution of the tumor automaton, where an immunotherapy round of treatment that took 250 steps was delivered to the tissue starting at step 300. 

\begin{figure}[h]
    \centering
    \includegraphics[width = \linewidth]{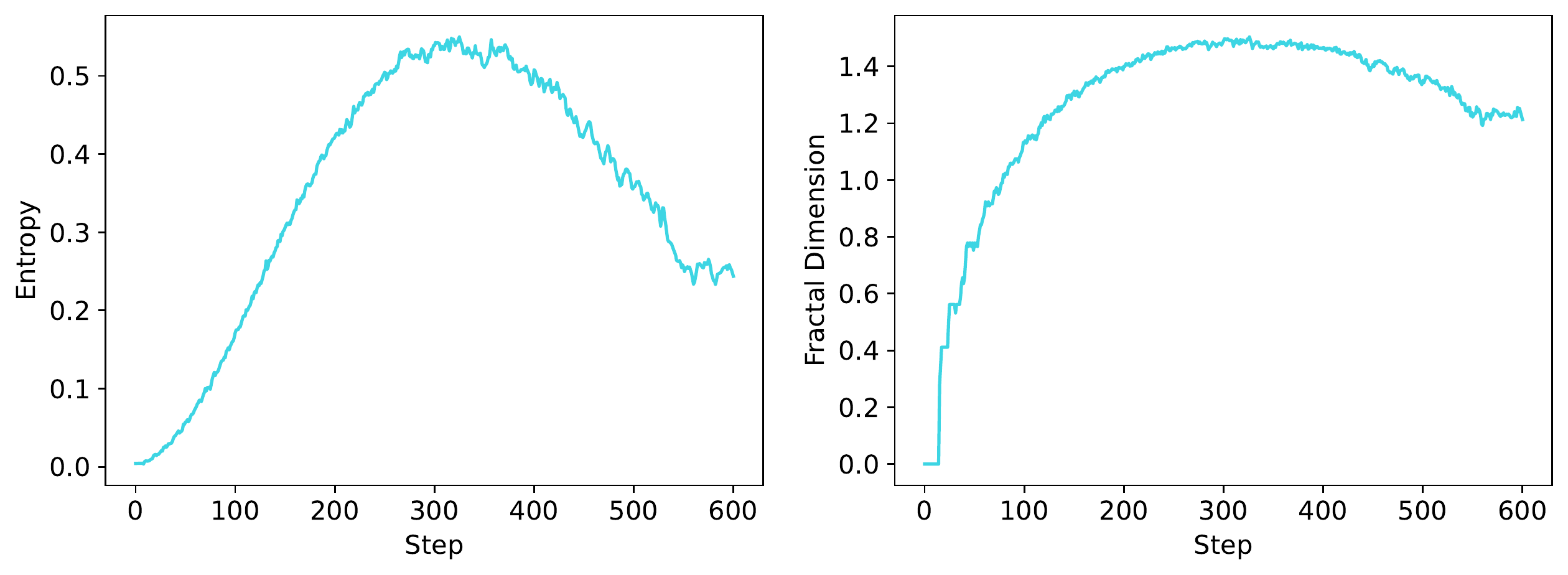}
    \caption{Entropy and fractal dimension time series computed in a case where immunotherapy was started at step 300.}
    \label{fig:immuno_entropy_frac}
\end{figure}

Corresponding time series of entropy and fractal dimension are shown in Figure \ref{fig:immuno_entropy_frac}. 

\section{Results}

Figures \ref{fig:automaton_state_evol} and \ref{fig:no_treatment_entropy_frac} show the results obtained with the avascular tumor growth model when no therapeutic interventions were administered. As the disease progresses, the tumor reaches a stationary final state, where cell count, entropy, and fractal dimension fluctuate around enduring values. This attractor of the cellular automaton's configuration, which represents the tumor, exhibits resilience and adaptative qualities toward possible external disturbances\cite{attractor_resilience}. 
Biologically this is very meaningful and correctly models resistance to treatment and recurrence phenomena, known traits of cancerous tumors.

The results obtained when radiotherapy was applied during the tumor evolution are shown in Figures \ref{fig:radio_automaton_state_evol} and \ref{fig:rad_treatment_entropy_frac}. It is observed that cell count, entropy, and fractal dimension decrease for some time after the radiation is delivered to the tissue, attempting to escape the tumor attractor. However, the re-incidence of cancer cell proliferation is observed, manifesting as three distal tumor sprouts, which suggests that the modeled intervention is insufficient to drive the system out of the tumor's potential well, so possibly multiple successive irradiation procedures would be required to eradicate the tumor. 

Chemotherapy results, shown in Figures \ref{fig:chemo_automaton_state_evol} and \ref{fig:chemo_entropy_frac}, yielded similar patterns to those obtained with radiotherapy. Namely, a temporary decrease of the analyzed variables was observed, followed by a recurrence in tumor growth. In addition, since chemotherapy spatially attacks cells following the drug's radial perfusion (in contrast to radiotherapy, where the targeting of the cells is more sparsely distributed), a more uniform reduction of the proliferating cells in the tumor's periphery was achieved.

Finally, results obtained for immunotherapy, presented in Figures \ref{fig:immuno_automaton_state_evol} and \ref{fig:immuno_entropy_frac}, showed similar patterns of decrease for state variables when the treatment was applied. In this case, however, the targeting of cancer cells was even more widely scattered than during radiotherapy, leaving multiple clusters of cells as the immune system attacked proliferating cells. This simulation resulted in slighter decreases for both entropy and fractal dimension, and it is an essential spatial aspect to consider since those cell groups could result in multiple separate tumor formations later on. 

\section{Discussion}

Comparison of time series for entropy and fractal dimension as the tumor grew and as each of the modeled treatments was delivered showed that while fractal dimension grows faster at initial stages of cancer development (greater slope), it responds slower and on a smaller magnitude than entropy to the applied therapeutic interventions. This phenomenon suggests that these biomarkers' sensitivity could vary depending on the tumor stage at which they are registered. Namely, the fractal dimension would be appropriate to detect initial changes in tissue as the tumor grows, while entropy could provide a useful measure to assess the effect of therapies on tissue. However, it is important to remark that the previous observations are limited by assumptions made in the automaton model. In addition, images generated with the model hold no direct relationship with any of the available imaging techniques used in a clinical setting. Thus, the sensitivity of the analyzed biomarkers could depend on the method used to record the tumor state. Nonetheless, this result highlights the importance of specifying the cancer stage whenever imaging biomarkers are investigated since their diagnostic and prognostic utility could be time-dependent.

A comparison of the examined interventions showed that both entropy and fractal dimension decreased more slowly whenever multiple scattered groups of cells were left due to the treatment (in our results, this occurred with immunotherapy). This finding suggests that time variations of biomarkers such as entropy and fractal dimension can reflect on relevant spatial evolution patterns of a complex system. In the particular case of a tumor, they could be associated with how efficient treatment is in disrupting communication and interactions of cancer cells in a tissular area. Both radiotherapy and chemotherapy showed a more homogeneous targeting of proliferating cells, obtaining relatively faster response curves for entropy and fractal dimension.

\section{Concluding remarks}

A cellular automata model for tumor growth and three treatment modalities (radiotherapy, chemotherapy, and immunotherapy) was developed in this work to study the evolution of 2 imaging biomarkers: Shannon's entropy and fractal dimension. Results from the model showed that adaptive and resilient properties of tumors toward treatment emerged naturally from the individual interactions of cancer cells. Namely, in all three treatments, cancer reoccurrence was observed at some point after discontinuing the interventions. Time series of entropy and fractal dimension obtained with a sequence of simulated tumor images showed that fractal dimension increased faster at the onset of cancer cell proliferation, while entropy exhibited the highest response to effects induced by cancer therapies. The previous result suggests that these biomarkers' sensitivity and prospective diagnostic utility could vary depending on the cancer stage and treatment conditions. Also, an imaging technique used to record the state of the tumor could play a critical role in their application for cancer evaluation purposes. A comparison among the simulated interventions revealed that entropy and fractal dimension decreased slower with therapies that left scattered, isolated, proliferating cells as they interacted with the tissue; this suggests that both quantities could reflect essential characteristics of the spatial targeting of cancer cells in therapies. 

\section*{Funding and conflict of interest}
The research leading to these results received funding from CONACYT-Mexico under Grant Agreement Fronteras 2019-217367.\\
The authors have no conflicts of interest to declare that are relevant to the content of this article.

\bibliographystyle{IEEEtran}
\bibliography{english_refs.bib}
\end{document}